\documentclass[10pt, final, conference, letterpaper, twocolumn, twoside]{IEEEtran}

\usepackage{cite}
\usepackage{graphicx}
\usepackage{subfigure}
\usepackage{amsmath}
\usepackage{amssymb}
\usepackage{epsfig}
\usepackage{times}
\usepackage{clrscode_mod}

\newtheorem{pavikl}{\em Lemma}
\newtheorem{pavikt}{\em Theorem}

\newcommand{\argmin}{\operatornamewithlimits{argmin}}

\begin{document}

\title{Enhancing Sensor Network Lifetime Using Interactive Communication}

\author{\IEEEauthorblockN{Samar Agnihotri and Pavan Nuggehalli}
\IEEEauthorblockA{Centre for Electronics Design and Technology, Indian Institute of Science, Bangalore - 560012, India.\\}
Email: \{samar, pavan\}@cedt.iisc.ernet.in%
}

\maketitle

\begin{abstract}
We are concerned with maximizing the lifetime of a data-gathering wireless sensor network consisting of set of nodes directly communicating with a base-station. We model this scenario as the $m$-message interactive communication between multiple correlated informants (sensor nodes) and a recipient (base-station). With this framework, we show that $m$-message interactive communication can indeed enhance network lifetime. Both worst-case and average-case performances are considered.
\end{abstract}

\section{Introduction}
\label{Intro}
Many future and extant sensor networks feature tiny sensor nodes with modest energy resources, processing power, and communication abilities. A key networking challenge is to devise protocols and architectures that can provide relatively long operational sensor network lifetimes, in spite of these limitations. Sensor nodes expend energy in sensing, computing, and communication. In this paper, we are concerned with reducing the energy cost of communication. We neglect the energy consumed by the nodes in sensing and computing because sensing costs are independent of the communication strategy being deployed and computing costs are often negligible compared to communication costs.

The energy expended by a sensor node in communication has two components: reception energy and transmission energy. The energy consumed in reception depends on the number of bits received and the per bit energy cost required to keep the receiver circuitry energized. The transmission energy depends on a number of factors such as transmit power levels, receiver sensitivity, channel state (including path loss due to distance and fading) and the kind of channel coding employed. In this paper, we assume that the data rates are low and that optimal channel coding is employed. Both these assumptions allow us to assume that the transmit power is linearly proportional to the data rate. Therefore, the communication energy is minimized by transmitting and receiving as few bits as possible.

In this work, based on some ideas from the theory of communication complexity, we propose a formalism to minimize the number of bits communicated in a single-hop sensor network, hence enhance the network lifetime. Assuming the correlation in sensor data, we model the communication between the base-station and sensor nodes as the $m$-message interactive communication between multiple correlated informants and a recipient, where at most $m$ messages are exchanged between a sensor node and the base-station. To the best of our knowledge, our work for the first time, employs this approach to estimate the sensor network lifetime.

\section{``Multiple Informants - Single Recipient'' Communication Complexity}
\label{sec1}
Let us consider a set of $N$ correlated informants \textit{interactively} communicating with a recipient. The objective of communication is that the recipient must learn about each informant's data with no probability of error, but the informants may or may not learn about the recipient's data.

Communication takes place over $N$ binary, error-free channels, where each channel connects an informant with the recipient. An informant and the recipient can communicate back and forth over the channel connecting them, but the informants cannot communicate directly with each other (although, they can communicate indirectly via the recipient). Each bit communicated over any channel, in either direction is counted. We want to estimate the number of bits exchanged in the worst and average cases for such communication scenarios.

Previously, \cite{090orlitsky, 091orlitsky, 092orlitsky, 094zhang, 097ahlswede} have attempted to bound the message complexity of ``single informant - single recipient'' communication. These efforts have shown that the number of bits exchanged depends on the number of messages exchanged, where the messages are finite sequences of bits and determined by agreed upon, deterministic protocol. However, only a little work \cite{arXiv0701048}, seems to have been done towards computing the message complexity of ``multiple correlated informants - single recipient'' interactive communication problem. The correlation among informants' data helps in reducing the problem of finding the optimal rates which minimize the communication complexity to the problem of finding an optimal schedule that minimizes the communication complexity \cite{arXiv0701048}. However, for an arbitrary model of correlation in informants' data, it is not straightforward to compute the optimal number of messages, which minimizes the number of bits exchanged. So, in this work, we develop the formalism to compute the number of bits exchanged for a given number $m$ of messages and an arbitrary model of correlation in informants' data.

In this work, we set $m = 2$ for three reasons. Firstly, it is shown in \cite{090orlitsky, 091orlitsky, 092orlitsky} that just two messages reduce the communication complexity exponentially compared to one message and at the same time, with just two messages, the number of bits exchanged is \textit{at worst} four times the optimal number of bits. Secondly, two is the minimum number of messages to show how the interaction helps in reducing communication complexity. Thirdly, in interactive communication, two messages give most pessimistic estimates of the worst and average case communication complexities.

In the rest of this section, we propose and illustrate our thesis to use the notions of ambiguity and information entropy to compute the worst and average case communication complexities, respectively.

\subsection{Ambiguity and Entropy}
\label{subsec:ambent}
We extend here the notions of \textit{ambiguity set} and \textit{ambiguity}, proposed in \cite{090orlitsky} and derive some of their properties.

Let $(X_1, X_2)$ be a random pair, $X_1 \in {\cal X}$ and $X_2 \in {\cal X}$, where ${\cal X}$ is discrete alphabet set\footnote{In general, $X_1 \in {\cal X}_1$ and $X_2 \in {\cal X}_2$, where ${\cal X}_1$ and ${\cal X}_1$ are discrete alphabet sets, with possibly different cardinalities. However, to keep the discussion simple, we assume henceforth that all the random variables take the values from same discrete alphabet ${\cal X}$, unless stated otherwise.}. Let $S_{X_1, X_2}$ denote the \textit{support set} of $(X_1, X_2)$. The \textit{support set} of $X_1$ is the set
\begin{equation*}
S_{X_1} \stackrel{\textrm{def}}{=} \{x_1: \mbox{ for some } x_2, (x_1, x_2) \in S_{X_1, X_2}\} \in {\cal X}_1,
\end{equation*}
of possible $X_1$ values. We also call $S_{X_1}$ \textit{ambiguity set} of $X_1$. The \textit{ambiguity} is defined as $\mu_{X_1} = |S_{X_1}|$ and it is same as the \textit{maximum ambiguity} $\widehat{\mu}_{X_1}$ of $X_1$.

The \textit{conditional ambiguity set} when random variable $X_2$ takes the value $x_2 \in S_{X_2}$ is
\begin{equation}
\label{eqn:ambiguity_set}
S_{X_1|X_2}(x_2) \stackrel{\textrm{def}}{=} \{x_1: (x_1, x_2) \in S_{X_1, X_2} \},
\end{equation}
the set of possible $X_1$ values when $X_2 = x_2$. The \textit{conditional ambiguity} in that case is
\begin{equation}
\label{eqn:ambiguity}
\mu_{X_1|X_2}(x_2) \stackrel{\textrm{def}}{=} |S_{X_1|X_2}(x_2)|,
\end{equation}
the number of possible $X_1$ values when $X_2=x_2$. The \textit{maximum conditional ambiguity} of $X_1$ is
\begin{equation}
\label{eqn:max_ambiguity}
\widehat{\mu}_{X_1|X_2} \stackrel{\textrm{def}}{=} \sup \{\mu_{X_1|X_2}(x_2): x_2 \in S_{X_2}\},
\end{equation}
the maximum number of $X_1$ values possible with any value that $X_2$ can take.

\begin{pavikl}
\label{lemma:conditionalAmbiguity}
$\mu_{X_1|X_2}(X_2 = x_2) \le \mu_{X_1}$, that is, conditioning reduces ambiguity.
\end{pavikl}
\begin{proof}
From the definitions of $\mu_{X_1}$ and $\mu_{X_1|X_2}(x_2)$, the proof is immediately obvious.
\end{proof}

Contrast this with a similar statement about entropy: $H(X_1|X_2 = x_2) \le H(X_1)$, which may or not hold always. Also, it follows from above lemma that $\widehat{\mu}_{X_1|X_2} \le \widehat{\mu}_{X_1}$.

\begin{pavikl}
\label{lemma:entropyAmbiguity}
$H(X_1|X_2 = x_2) \le \log{\mu_{X_1|X_2}(X_2=x_2)}$.
\end{pavikl}
\begin{proof}
The proof follows from the definitions of $H(X_1|X_2 = x_2)$ and $\log{\mu_{X_1|X_2}(X_2=x_2)}$. Note that equality is achieved in the statement of the lemma only when $p(X_1/X_2=x_2)$ is uniformly distributed.
\end{proof}

Taking the expectation of the both side of the inequality in Lemma \ref{lemma:entropyAmbiguity} with respect to $p(x_2)$ gives:
\begin{equation}
H(X_1|X_2) \le \sum_{x_2 \in {\cal X}_2} \!\!\! p(X_2 = x_2) \log{\mu_{X_1|X_2}(x_2)}.
\end{equation}

Let us define \textit{average ambiguity} of $X_1$ as
\begin{equation}
\label{eqn:av_ambiguity}
\overline{\mu}_{X_1|X_2} \stackrel{\textrm{def}}{=} \sum_{x_2 \in S_{X_2}} \!\!\!\! p(X_2 = x_2) \mu_{X_1|X_2}(x_2),
\end{equation}
the average number of $X_1$ values possible with all $X_2$ values. It is obvious from Lemma \ref{lemma:conditionalAmbiguity} that $\overline{\mu}_{X_1|X_2} \le \overline{\mu}_{X_1}$.

These definitions are naturally extended to $N$ random variables $X_1, \ldots, X_N$. For example:
\begin{equation*}
\mu_{X_1|X_2, \ldots, X_N}(x_2, \ldots, x_N) \stackrel{\textrm{def}}{=} |S_{X_1|X_2, \ldots, X_N}(x_2, \ldots, x_N)|.
\end{equation*}

\subsection{Notation}
\label{subsec:notation}
In this sub-section, we introduce the notation that will be frequently used in the rest of this paper.
\begin{description}
\item [$S$:] the set of $N$ informants.
\item [$\cal X$:] finite, discrete alphabet set. $|{\cal X}| = n$.
\item [$\Pi$:] the set of all $N!$ schedules to poll $N$ informants.
\item [$\pi(i)$:] the informant polled at $i^{\textrm{th}}$ slot in the schedule $\pi$.
\item [$A_{\pi(i)}$:] the set $\{\pi(1), \ldots, \pi(i-1)\}$ of informants who have already communicated their data to the recipient before the $i^{\textrm{th}}$ informant in the schedule $\pi$.
\item [$X_{\pi(1)}, \ldots, X_{\pi(i - 1)}$:] \hspace{0.8in} is denoted as $X_{A_{\pi(i)}}$.
\item [$p(X_{\pi(1)} = x_{\pi(1)}, \ldots, X_{\pi(i - 1)} = x_{\pi(i - 1)})$:] \hspace{2.01in} is denoted as $p(x_{A_{\pi(i)}})$.
\item [$p(X_{\pi(i)} = x_{\pi(i)} | X_{\pi(1)} = x_{\pi(1)}, \ldots, X_{\pi(i - 1)} = x_{\pi(i - 1)})$:] \hspace{2.81in} is denoted as $p(x_{\pi(i)}|x_{A_{\pi(i)}})$.
\end{description}

\subsection{``Single Informant - Single Recipient'' communication}
\label{subsec:singleInformantsingleRecipient}
Let us consider two message interactive communication between two persons $P_{\cal X}$ and $P_{\cal Y}$. $P_{\cal X}$ observes the random variable $X_1 \in {\cal X}$ and $P_{\cal Y}$ observes a possibly dependent random variable $X_2 \in {\cal X}$. Let us assume that only $P_{\cal Y}$ knows the joint distribution $p(x_1, x_2)$. In the worst-case, $P_{\cal Y}$ needs to send $\max(1, \lceil \log \log \widehat{\mu}_{X_1|X_2} \rceil)$ bits to $P_{\cal X}$  to help it send its information in $\lceil \log \widehat{\mu}_{X_1|X_2} \rceil$ bits to let $P_{\cal Y}$ learn about $X_1$. Similarly, on average, $P_{\cal Y}$ needs to send \textit{at least} $H(X_1|X_2)$ bits to $P_{\cal X}$ to help it send its information in $H(X_1|X_2)$ bits to let $P_{\cal Y}$ learn about $X_1$. However, we soon show that in the optimal communication protocol the recipient $P_{\cal Y}$ needs to send many more bits than the number of bits given above.

In the following, we generalize this discussion to ``$N$ correlated informants - single recipient'' communication problem and show that the interactive communication between the recipient and informants using \textit{prefix-free} messages, reduces this problem to a scheduling problem where the order, in which $N$ ``single recipient - single informant'' communication problems are solved, is to be computed. We compute the worst and average case communication complexities and give optimal communication protocols, with at most two messages exchanged between any informant and the recipient. Assume that the joint probability distribution $p(x_1, \ldots, x_N), x_i \in {\cal X}$, of informants' data is \textit{only} known to the recipient. Contrast this with the communication scenarios considered in \cite{090orlitsky, 091orlitsky, 092orlitsky}, where both, the recipient and informant know the joint distribution. The immediate consequence of this difference, as shown below, is that both in the worst-case and average-case, the recipient needs to communicate more bits in our setup.

\subsection{Worst-case communication complexity}
\label{subsec:worstcase}
Let us consider a communication schedule $\pi \in \Pi$. Let us assume that the informants $\pi(1), \ldots, \pi(i-1)$ have already communicated their data to the recipient. The conditional ambiguity set of the recipient of informant $\pi(i)$'s data is $S_{X_{\pi(i)}|X_{A_{\pi(i)}}}$, with maximum ambiguity $\widehat{\mu}_{X_{\pi(i)}|X_{A_{\pi(i)}}} \le n$. In the worst-case, the recipient requires $\lceil \log \widehat{\mu}_{X_{\pi(i)}|X_{A_{\pi(i)}}} \rceil$ bits to learn unambiguously about the informant $\pi(i)$'s data. So, it is both necessary and sufficient that the informant $\pi(i)$ sends $\lceil \log \widehat{\mu}_{X_{\pi(i)}|X_{A_{\pi(i)}}} \rceil$ bits to the recipient. However, to help the informant $\pi(i)$ send its information in just these many bits, the recipient informs it in $\lceil \log n \rceil$ bits about the index of each element of $S_{X_{\pi(i)}|X_{A_{\pi(i)}}}$. Then, the informant $\pi(i)$ constructs the prefix-free codes corresponding to those data values and sends the code corresponding to its actual data value to the recipient in $\lceil \log \widehat{\mu}_{X_{\pi(i)}|X_{A_{\pi(i)}}} \rceil$ bits.

Following this protocol to poll all the informants, the total number of bits transmitted by recipient under schedule $\pi$, is
\begin{eqnarray}
\label{eqn:recipientWorstCase}
\widehat{R}_{\pi} \!\!\!\!\! & = & \!\!\!\!\! \sum_{i=1}^N \widehat{B}_{R, \pi(i)} \\
& = & \!\!\!\!\! \sum_{i=1}^N \widehat{\mu}_{X_{\pi(i)}|X_{A_{\pi(i)}}} \lceil \log n \rceil. \nonumber
\end{eqnarray}
The total number of bits transmitted by the informant $\pi(i)$ is
\begin{equation}
\label{eqn:informantsWorstCase}
\widehat{I}_{\pi(i), R} = \lceil \log \widehat{\mu}_{X_{\pi(i)}|X_{A_{\pi(i)}}} \rceil.
\end{equation}
We are interested in finding the schedule $\pi^*$ that solves:
\begin{equation}
\label{eqn:worstCaseProblem}
\pi^* = \argmin_{\pi \in \Pi} \max_{i = 1, \ldots, N}\widehat{I}_{\pi(i), R}.
\end{equation}

The $\min\max$ nature of the problem in \eqref{eqn:worstCaseProblem} ensures that the \textit{Minimum Cost Next (MCN)} algorithm described below computes the optimal schedule for the problem in \eqref{eqn:worstCaseProblem}.

\hspace{-0.5cm}\hrulefill

\hspace{-0.25cm}{\textbf{Algorithm:} MCN}

\vspace{-0.2cm}\hspace{-0.5cm}\hrulefill
\begin{codebox}
\li Initialization: $k = 1$, $A_{\pi^{MCN}(k)} = \phi$.
\li \While $(k \leq N)$
\li $\pi^{MCN}(k) = \argmin_{i \in S-A_{\pi^{MCN}(k)}} \widehat{I}_{i, R}$. \label{li:next-node}
\li $A_{\pi^{MCN}(k+1)} = A_{\pi^{MCN}(k)} \cup \pi^{MCN}(k)$.
\li $k = k + 1$.
    \End
\end{codebox}
\vspace{-0.2cm}\hrulefill

\begin{pavikl}
\label{lemma:worstmcnMax}
\textit{MCN} schedule solves \eqref{eqn:worstCaseProblem}.
\end{pavikl}
\begin{proof}
Omitted for brevity.
\end{proof}

\subsection{Average-case communication complexity}
\label{subsec:avcase}
In the scenarios that we are interested in, the marginal and conditional probabilities are available only to the recipient, but not to the individual informants. We propose a protocol for interactive communication between the recipient and an informant, where the recipient constructs the optimal source code for each element of its ambiguity set of an informant's data, communicates those to the informant, which responds by communicating the code corresponding to its information.

Let us consider a schedule $\pi$ and assume that the informants $\pi(1), \ldots, \pi(i-1)$ have communicated their data to the recipient. The conditional ambiguity set of the recipient of informant $\pi(i)$'s data is $S_{X_{\pi(i)}|X_{A_{\pi(i)}}}$, with ambiguity ${\mu}_{X_{\pi(i)}|X_{A_{\pi(i)}}} \le n$. The recipient computes $p(x_{\pi(i)}|x_{A_{\pi(i)}})$ and the corresponding Huffman code for the elements of $S_{X_{\pi(i)}|X_{A_{\pi(i)}}}$. Then, in $\lceil \log n \rceil$ bits it conveys to the informant $\pi(i)$ the index of each element of $S_{X_{\pi(i)}|X_{A_{\pi(i)}}}$ and in $l_j, j = 1, \ldots, \mu_{X_{\pi(i)}|X_{A_{\pi(i)}}}$ bits the corresponding Huffman code.

Following this protocol to poll all the informants, the total number of bits transmitted by recipient under schedule $\pi$, is
\begin{eqnarray}
\overline{R}_{\pi} \!\!\!\!\!\! & = & \!\!\!\!\!\! \sum_{i=1}^N \overline{B}_{R, \pi(i)} \label{eqn:recipientAvCase} \\
& = & \!\!\!\!\!\! \sum_{i=1}^N \!\!\!\! \sum_{\stackrel{x_{A_{\pi(i)}}}{\in S_{X_{A_{\pi(i)}}}}} \!\!\!\!\!\!\! p(x_{A_{\pi(i)}}\!) \! \bigg( \! \mu_{X_{\pi(i)}|X_{A_{\pi(i)}}} \!\! \lceil \log n \rceil \! + \!\!\!\!\!\! \!\!\!\!\!\! \sum_{\stackrel{x_{\pi(i)}}{\in S_{X_{\pi(i)}|X_{A_{\pi(i)}}}}} \!\!\!\!\!\! \!\!\!\!\!\! \! l_{x_{\pi(i)}|x_{A_{\pi(i)}}} \!\! \bigg) \nonumber \\
& = & \!\!\!\!\!\! \sum_{i=1}^N \!\! \bigg( \overline{\mu}_{X_{\pi(i)}|X_{A_{\pi(i)}}} \lceil \log n \rceil + \!\!\!\!\!\! \!\!\!\! \sum_{\stackrel{x_{\pi(1)}, \ldots, x_{\pi(i)}}{\in S_{X_{\pi(1)}, \ldots, X_{\pi(i)}}}} \!\!\!\!\!\! \!\!\!\!\!\! \!\! p(x_{A_{\pi(i)}}) l_{x_{\pi(i)}|x_{A_{\pi(i)}}} \!\!\bigg). \nonumber
\end{eqnarray}
The total number of bits transmitted by the informant $\pi(i)$ is
\begin{eqnarray}
\overline{I}_{\pi(i), R} \!\!\!\!\! & = & \!\!\!\!\! \sum_{\stackrel{x_{A_{\pi(i)}}}{\in S_{X_{A_{\pi(i)}}}}} \!\!\!\!\!\! p(x_{A_{\pi(i)}}) \!\!\!\! \sum_{\stackrel{x_{\pi(i)}}{\in S_{X_{\pi(i)}|X_{A_{\pi(i)}}}}} \!\!\!\!\!\! \!\!\!\!\!\! p(x_{\pi(i)}|x_{A_{\pi(i)}}) l_{x_{\pi(i)}|x_{A_{\pi(i)}}} \nonumber \\
& = & \!\!\!\!\! \sum_{\stackrel{x_{\pi(1)}, \ldots, x_{\pi(i)}}{\in S_{X_{\pi(1)}, \ldots, X_{\pi(i)}}}} \!\!\!\!\!\! p(x_{\pi(1)}, \ldots, x_{\pi(i)}) l_{x_{\pi(i)}|x_{A_{\pi(i)}}}. \label{eqn:informantsAvCase}
\end{eqnarray}
The objective is to find schedule $\pi^*$ that solves:
\begin{equation}
\label{eqn:avCaseProblem}
\pi^* = \argmin_{\pi \in \Pi} \max_{i = 1, \ldots, N}\overline{I}_{\pi(i), R}.
\end{equation}

\begin{pavikl}
\label{lemma:avmcnMax}
\textit{MCN} schedule solves \eqref{eqn:avCaseProblem}.
\end{pavikl}
\begin{proof}
Omitted for brevity.
\end{proof}

Note that we can easily design sophisticated two message communication protocols that reduce $\widehat{R}_{\pi}$ and $\overline{R}_{\pi}$ to their minimum possible values, yet solve \eqref{eqn:worstCaseProblem} and \eqref{eqn:avCaseProblem}, respectively. However, here we omit the details of such protocols for the sake of brevity.

\section{System Model}
\label{systemModel}
We consider a network of $N$ battery operated sensor nodes strewn in a coverage area. The nodes are assumed to interactively communicate with the base-station in a single hop. Sensor node $k, k \in \{1, \ldots, N\}$ has $E_k$ units of energy and the base-station has $E_{BS}$ units of energy. The wireless channel between sensor $k$ and the base-station is described by a symmetrical path loss $d_k$, which captures various channel effects and is assumed to be constant. This is reasonable for static networks and also for the scenarios where the path loss varies slowly and can be accurately tracked.

The network operates in a time-division multiple access (TDMA) mode. Time is divided into slots and in each slot, every sensor communicates its data to the base-station. Let us assume that the sensor data at every time slot is described by a random vector $(X_1, \ldots, X_N) \sim p(x_1, \ldots, x_N), x_i \in {\cal X}$. This distribution is \textit{only} known to the base-station. We assume the spatial correlation in the sensor data and ignore temporal correlation, as it can easily be incorporated in our work for data sources satisfying the Asymptotic Equipartition Property.

We assume static scheduling, that is the base-station uses the same sensor polling schedule in every time slot, until the network dies. The worst-case lifetime of a sensor node (base-station) under schedule $\pi \in \Pi$ is defined as the ratio of its total energy and its worst-case energy expenditure in a slot, under schedule $\pi$. However, as argued in Introduction, it is only the communication energy expenditure that we are here concerned with. The average lifetime of a sensor node (base-station) is similarly defined. We define network lifetime as the time until the first sensor node or the base-station runs out of the energy. This definition has the benefit of being simple, practical, and popular \cite{Tassiulas} and as shown below, provides a $\max\min$ formulation of the network lifetime in terms of the lifetimes of the sensor nodes and the base-station.

To model the transmit energy consumption at the base-station and the sensor nodes, we assume that transmission rate is linearly proportional to signal power. This assumption is motivated by Shannon's AWGN capacity formula which is approximately linear for low data rates. So, a node $k$ under schedule $\pi$ expends $B_{\pi(k)} d_k$ units of energy to transmit $B_{\pi(k)}$ units of information. Let $E_r$ denote the energy cost of receiving one bit of information. For simplicity, let us assume that it is same for both the base-station and the sensor nodes.

The general problem is to find the optimal rates (the number of bits to transmit), which maximize network lifetime. However, the optimal rate-allocation is constrained to lie within the Slepian-Wolf achievable rate region. This makes the problem computationally challenging. We simplify the problem by introducing the notion of \textit{instantaneous decoding} \cite{wowmom05} and thus reduce the optimal rate allocation problem to computing the optimal scheduling order, albeit at some loss of optimality. This loss of optimality occurs because, in general, turning a multiple-access channel into an array of orthogonal channels by using a suitable MAC protocol (TDMA in our case) is well-known to be a suboptimal strategy, in the sense that the set of rates that are achievable with orthogonal access is strictly contained in the Ahlswede-Liao capacity region \cite{Cover_book}.

\section{Maximizing Sensor Network Lifetime}
\label{sensorNet_appl}
Let us assume that the interaction between the base-station and the sensor nodes is not allowed. Then in the worst-case, every node sends $\lceil \log n \rceil$ bits to the base-station to convey its information. However, if every node knows $p(x_1, \ldots, x_N)$ and the data of all other nodes, then it only needs to send the bits describing its data conditioned on the data of the nodes already polled \cite{Cristescu}. In the real single-hop sensor networks, neither it is possible that every node knows about all other nodes' data, given the limited communication capabilities of the sensor nodes; nor it is desired that the sensor nodes perform such computationally intense processing, given their limited computational and energy capabilities.

However, if we allow the interaction between the base-station and sensor nodes, then the nodes can still send less than $\lceil \log n \rceil$ bits, yet avoid above issues. In fact, this is precisely the ``multiple correlated informants - single recipient'' communication problem of section \ref{sec1}. Using the results derived there and identifying the recipient as the base-station and informants as the sensor nodes, in the following, we attempt to maximize the worst and average case lifetimes of the single-hop sensor networks, for the given model of energy consumption and spatial correlation in the sensor data. The base-station and a sensor node interactively communicate by exchanging at most two messages. To estimate the worst and average case lifetimes of the sensor networks, we use the protocols in \ref{subsec:worstcase} and \ref{subsec:avcase}, respectively for the base-station and sensor nodes communication. With these protocols, the maximum number of bits transmitted by any sensor node is minimized and the base-station carries most of the burden of computation and communication in the network. This is reasonable in the scenarios where the base-station is computationally and energy-wise more capable than the sensor nodes. Still, it may not be infinitely more capable. So, in the network lifetime estimation problem, we consider the total communication (transmission and reception) energy expenditure at every sensor node as well as the base-station.

\subsection{Worst-case Network Lifetime}
\label{subsec:worstNetLife}
Let $\widehat{E}_{BS, \pi(i)}$ denote the energy that the base-station spends in communicating with node $\pi(i)$ in the worst-case, that is, it denotes the energy that the base-station spends in transmitting and receiving the bits from node $\pi(i)$, in the worst-case. So,
\begin{equation}
\widehat{E}_{BS, \pi(i)} = \widehat{B}_{BS, \pi(i)} d_i + \widehat{I}_{\pi(i), BS} E_r.
\end{equation}
Similarly, let $\widehat{E}_{\pi(i), BS}$ denote the energy that the node $\pi(i)$ spends in communicating with the base-station. So,
\begin{equation}
\widehat{E}_{\pi(i), BS} = \widehat{I}_{\pi(i), BS} d_i + \widehat{B}_{BS, \pi(i)} E_r.
\end{equation}
On substituting for $\widehat{B}_{BS, \pi(i)}$ and $\widehat{I}_{\pi(i), BS}$ from \eqref{eqn:recipientWorstCase} and \eqref{eqn:informantsWorstCase}, respectively, we have
\begin{eqnarray}
\widehat{E}_{BS, \pi(i)} \! - \widehat{E}_{\pi(i), BS} \!\!\!\!\! & = & \!\!\!\!\! \big(\widehat{\mu}_{X_{\pi(i)}|X_{A_{\pi(i)}}} \lceil \log n \rceil \label{eqn:energyDifference} \\
& & + \lceil \log \log \widehat{\mu}_{X_{\pi(i)}|X_{A_{\pi(i)}}} \rceil \nonumber \\
& & - \lceil \log \widehat{\mu}_{X_{\pi(i)}|X_{A_{\pi(i)}}} \rceil\big) (d_i - E_r). \nonumber
\end{eqnarray}
Assuming $d_i \ge E_r$, this implies that $\widehat{E}_{BS, \pi(i)} - \widehat{E}_{\pi(i), BS} \ge 0$, that is, the base-station spends more energy in communicating with node $\pi(i)$ than vice versa.

Given our definitions of the sensor node, the base-station, and the network lifetimes, the worst-case lifetime $\widehat{L}$ of the network is the solution to the following optimization problem
\begin{eqnarray}
\widehat{L} = \max_{\pi \in \Pi} \, \min \Big(\frac{E_{BS}}{\sum_{i=1}^N \widehat{E}_{BS, \pi(i)}} \,, \min_{i = 1, \ldots, N} \frac{E_{\pi(i)}}{\widehat{E}_{\pi(i), BS}}\Big) \label{eqn:worstLifetime1}, \\
\widehat{L}^{-1} = \min_{\pi \in \Pi} \, \max \Big(\frac{\sum_{i=1}^N \widehat{E}_{BS, \pi(i)}}{E_{BS}} \,, \max_{i = 1, \ldots, N} \frac{\widehat{E}_{\pi(i), BS}}{E_{\pi(i)}}\Big). \label{eqn:worstLifetime2}
\end{eqnarray}

Before we discuss the nature of the general solution to this problem, let us consider its two special cases.

\textit{Case 1:} Let $E_{BS} = E_1 = \ldots = E_N = E$. This is so when $N+1$ identical sensors form a sensor cluster and one of those sensor nodes, is also chosen as the clusterhead. Then, the problem in \eqref{eqn:worstLifetime2} reduces to
\begin{equation*}
\widehat{L}^{-1} = \frac{1}{E}\min_{\pi \in \Pi} \, \max \Big(\sum_{i=1}^N \widehat{E}_{BS, \pi(i)} \,, \max_{i = 1, \ldots, N} \widehat{E}_{\pi(i), BS}\Big).
\end{equation*}
However, $\sum_{i=1}^N \widehat{E}_{BS, \pi(i)} \ge \max_{i = 1, \ldots, N} \widehat{E}_{\pi(i), BS}$ as implied by \eqref{eqn:energyDifference}, so we have
\begin{equation}
\label{eqn:case1}
\widehat{L}^{-1} = \frac{1}{E}\min_{\pi \in \Pi} \sum_{i=1}^N \widehat{E}_{BS, \pi(i)}.
\end{equation}
Following lemma computes the optimal solution for \eqref{eqn:case1}.

\begin{pavikl}
\label{lemma:mcnSum}
\textit{MCN} schedule solves
\begin{equation}
\label{eqn:BSlifetime}
\min_{\pi \in \Pi} \frac{\sum_{i=1}^N \widehat{E}_{BS, \pi(i)}}{E_{BS}}.
\end{equation}
\end{pavikl}
\begin{proof}
Changing the line~\ref{li:next-node} of the \textit{MCN} algorithm in \ref{subsec:worstcase} to $\pi^{MCN}(k)=\argmin_{i \in S-A} \sum_{j \in A \cup i} \widehat{E}_{BS, j}$, we obtain a version of the \textit{MCN} algorithm that solves \eqref{eqn:BSlifetime}. However, we omit the details for the sake of brevity.
\end{proof}

\textit{Case 2:} Let $E_1 = \ldots = E_N = E$, but $E_{BS} \gg E$. This is so when the base-station is \textit{infinitely} more capable than any of the identical sensor nodes. Then, \eqref{eqn:worstLifetime2} reduces to
\begin{eqnarray}
\widehat{L}^{-1} \!\!\!\!\! & = & \!\!\!\!\! \frac{1}{E_{BS}}\min_{\pi \in \Pi} \max \! \Big(\sum_{i=1}^N \widehat{E}_{BS, \pi(i)}, \frac{E_{BS}}{E} \max_{i = 1, \ldots, N} \widehat{E}_{\pi(i), BS} \Big) \nonumber \\
 & = & \frac{1}{E}\min_{\pi \in \Pi} \max_{i = 1, \ldots, N} \widehat{E}_{\pi(i), BS}, \mbox{ for } E_{BS} \gg E. \label{eqn:case2}
\end{eqnarray}
Lemma below computes the optimal lifetime in \eqref{eqn:case2}.

\begin{pavikl}
\label{lemma:mcnMax}
\textit{MCN} schedule solves
\begin{equation}
\label{eqn:sensorslifetime}
\min_{\pi \in \Pi} \max_{i = 1, \ldots, N} \frac{\widehat{E}_{\pi(i), BS}}{E_{\pi(i)}}.
\end{equation}
\end{pavikl}
\begin{proof}
Changing the line~\ref{li:next-node} of the \textit{MCN} algorithm in \ref{subsec:worstcase} to $\pi^{MCN}(k)=\argmin_{i \in S-A} \frac{\widehat{E}_{i, BS}}{E_{i}}$, we obtain a version of the \textit{MCN} algorithm that solves \eqref{eqn:sensorslifetime}. However, once more, we omit the details for the sake of brevity.
\end{proof}

The general problem in \eqref{eqn:worstLifetime1} or equivalently in \eqref{eqn:worstLifetime2} can be solved as follows. Let us define
\begin{eqnarray}
\pi_{sum} & = & \argmin_{\pi \in \Pi} \frac{\sum_{i=1}^N \widehat{E}_{BS, \pi(i)}}{E_{BS}}, \label{sumMCN}\\
\pi_{max} & = & \argmin_{\pi \in \Pi} \max_{i = 1, \ldots, N} \frac{\widehat{E}_{\pi(i), BS}}{E_{\pi(i)}}. \label{maxMCN}
\end{eqnarray}
It follows from Lemmas \ref{lemma:mcnSum} and \ref{lemma:mcnMax} that $\pi_{sum}$ and $\pi_{max}$ are the \textit{MCN} schedules which solve \eqref{sumMCN} and \eqref{maxMCN}, respectively. Let $S^{MCN} = \{\pi_{sum}, \pi_{max}\}$. Then, \eqref{eqn:worstLifetime2} reduces to:
\begin{equation}
\label{eqn:optimalSoln}
\widehat{L}^{-1} = \min_{\pi \in S^{MCN}} \, \max \Big(\frac{\sum_{i=1}^N \widehat{E}_{BS, \pi(i)}}{E_{BS}} \,, \max_{i = 1, \ldots, N} \frac{\widehat{E}_{\pi(i), BS}}{E_{\pi(i)}}\Big).
\end{equation}

\begin{pavikt}
\label{thrm:optimalSoln}
$\widehat{L}^{-1}$ in \eqref{eqn:optimalSoln} is optimal.
\end{pavikt}
\begin{proof}
Omitted for brevity.
\end{proof}

\subsection{Average Network Lifetime}
\label{subec:avNetLife}
Let $\overline{E}_{BS, \pi(i)}$ denote the energy that the base-station spends in communicating with node $\pi(i)$, on average. So,
\begin{equation}
\overline{E}_{BS, \pi(i)} = \overline{B}_{BS, \pi(i)} d_i + \overline{L}_{X_{\pi(i)}|X_{A_{\pi(i)}}} E_r.
\end{equation}

Similarly, let $\overline{E}_{\pi(i), BS}$ denote the energy that the node $\pi(i)$ spends in communicating with base-station. So,
\begin{equation}
\overline{E}_{\pi(i), BS} = \overline{L}_{X_{\pi(i)}|X_{A_{\pi(i)}}} d_i + \overline{B}_{BS, \pi(i)} E_r.
\end{equation}

Then, the average-case lifetime $\overline{L}$ of the network is the solution to the following optimization problem
\begin{equation}
\label{eqn:averageLifetime}
\overline{L} = \max_{\pi \in \Pi} \, \min \Big(\frac{E_{BS}}{\sum_{i=1}^N \overline{E}_{BS, \pi(i)}} \,, \min_{i = 1, \ldots, N} \frac{E_{\pi(i)}}{\overline{E}_{\pi(i), BS}}\Big).
\end{equation}

Identifying conditional ambiguity in section \ref{subsec:worstNetLife} as the conditional entropy and then following the same reasoning, all the discussion and results there hold true here as well.

\section{Conclusions and Future Work}
\label{conclusions}
We computed the worst and average case communication complexities for ``multiple correlated informants - single recipient'' communication, assuming that at most two messages are exchanged between an informant and the recipient and only the recipient knows the joint distribution of informants' data. Then, we applied these results to estimate the worst and average case lifetimes of the sensor networks.

However, there are other interesting variations of the interactive communication problem considered here and their applications to the sensor networks, such as the one where each informant also knows of the joint distribution of all informants' data, and the ones where $\widehat{R}_{\pi}$ and $\overline{R}_{\pi}$ are also included in the optimization problems \eqref{eqn:worstCaseProblem} and \eqref{eqn:avCaseProblem}, respectively. At present, we are working on such problems.

\end{document}